# Power System Decarbonization: Impacts of Energy Storage Duration and Interannual Renewables Variability

Mehdi Jafari[1], Magnus Korpas[2], and Audun Botterud[1]

[1]Laboratory for Information and Decision Systems (LIDS), MIT, Cambridge, MA, USA
[2]Norwegian University of Science and Technology (NTNU), Trondheim, Norway

*Abstract:*

Decarbonization of the electricity sector is one of the major measures in slowing down the pace of climate change. In this paper, we analyze the impacts of energy storage systems (ESS) and year-to-year variability and uncertainty in the hourly profiles of variable renewable energy (VRE) on power system decarbonization in 2050. We perform this analysis through capacity expansion optimization based on technology cost projections and $CO_2$ emission restrictions based on 11 years of wind, solar, and load data variations in Italy's power system, with a particular focus on how ESS changes the optimal generation portfolio and system performance. We also explore the impact of ESS duration on the renewables' adoption and system costs. To quantify the impact of VRE variability in different years, we present a comparative analysis of capacity expansion optimization based on multiple-year and single-year data. Our results indicate a high RES penetration even in the absence of decarbonization policies, due to expected declines in future technology costs. In the transition to a zero $CO_2$ emissions system, carbon capture and storage (CCS) plays a minor role due to its carbon capture efficiency, which is less than 100%. ESS investments contribute to lower total system costs by replacing more expensive flexibility resources. However, the value of ESS changes by its duration, with longer ESS durations having lower marginal value per added kWh storage capacity. Variability in VRE profiles leads to substantial variation in the system's configuration and energy cost, depending on what year is used in the optimization. Decision making based on single-year data can therefore lead to substantial increases in the systems' operational costs in other years due to increased probability of capacity shortages and load curtailments. In contrast, optimizing over multiple VRE and load years provides a more robust and cost-effective generation expansion strategy.

*Keywords:*

Renewable energy sources, energy storage, decarbonization, data uncertainty, capacity expansion

## 1. Introduction

Expansion of renewable energy has become the main solution for transitioning towards a low-carbon energy supply and to address global climate change reported by Intergovernmental Panel on Climate Change (IPCC) in 2018 [1], [2]. In the electricity and heat sectors, which account for 25% of overall greenhouse gas (GHG) emissions, integration of renewable resources reduce the GHG emissions [3].

Decarbonization of electricity system can be achieved by integration of renewable energy sources (RES) such as hydro, solar and wind and biofueled power plants [4]. In [5], the authors have assessed the potential of solar and wind energy in the realization of low-carbon energy systems, comparing advanced modeling approaches to illustrate the contribution of variable renewable energy (VRE) in different decarbonization scenarios. The authors in [6] show that decarbonization in the Greek energy system is possible through wide scale penetration of RES, if transmission system expansion and flexibility requirements are provided. However, many studies suggest that deep penetration of RES is not viable without energy storage systems (ESS), due to challenges such as intermittency of VRE sources, required flexibility in electricity supply, and power quality issues [7]–[10].

With increased interest in ESS, their performance in power system applications are extensively studied in the literature, e.g. [11]–[13]. As shown in [14], the value of solar and wind energy can be increased when combined with ESS, however this added value depends on the storage technology and its costs. Also, it is important to consider the energy storage size in different renewable energy systems, as the sizing criteria change depending on the type of renewable energy technology [15]. Different sizes and durations of ESS can also change the curtailment rate of different renewable technologies, although VRE curtailment is as a function of their penetration level [16]. In a more detailed study in [17], the authors have explored the value of ESS and system configurations with nuclear and renewable resources under different $CO_2$ emission constraints.

They concluded that size and duration of ESS and stricter $CO_2$ constraints has significant effect on the value of ESS. Moreover, substantial reductions in ESS costs are required for a large deployment of ESS to be economically viable. The duration of ESS has recently received increasing attention in the literature [18]–[21]. For instance, the U.S. Department of Energy has announced its duration addition to electricity storage (DAYS) program [18] seeking low cost solutions for ESS technologies with durations between 10 and 100 hours. Flow batteries [21], hydrogen [20], compressed air energy storage [22] and pumped hydro storage are examples of long duration ESS technologies with different characteristics and costs. However, it is not clear if these long duration ESS are economically viable in the power system operation and how much value they add to the system.

As indicated in the above-mentioned studies, low-emission power systems can be achieved with high penetration of renewables. However, there is another important issue that is less explored in the studies, namely the annual variations of the VRE outputs and loads and how these may influence the generation expansion plan. The VRE output pattern and hourly demand change from year to year. However, most of the capacity expansion studies are conducted based on a single year of renewables' data. Decision-making based on single-year data can lead to suboptimal operation in other years and increased system costs due to insufficient installed capacity and unserved demand. The authors in [23] have studied the impact of VRE variations on the installed capacity of flexible units (gas power plants and energy storage) and zonal VRE capacities under 50 and 80% RES penetration levels. In [24], the European system's operation is explored under year-to-year variations of the wind and solar energy. These studies show that the variation in VRE output can significantly change the system's costs. However, there are aspects that should be further explored such as zero-carbon system configuration and requirements, low emission flexibility plants (CCS and biofuel) and impact of VRE uncertainty on the installed capacity of all generation technologies.

Building on the previous studies and to address the mentioned shortcomings, we present a quantitative analysis of the pathway towards decarbonization in 2050 under different ESS, VRE and load uncertainty scenarios. Using Italy's power system as the candidate system, we conduct an analysis of the least-cost planning and operation and investigate the optimal technology configurations and costs of the system. We use an integrated investment and operation optimization model instead of soft-linking optimization of the planning and operation [23], [25]–[27]. One benefit of the integrated model is to avoid suboptimal operation and increased costs, which may happen under soft-linking as the capacity expansion does not consider operations in detail. We focus on the following three dimensions: (1) capacity expansion under different $CO_2$ emission constraints with and without battery energy storage deployment across 11 years of VRE availability and load data, (2) different ESS energy capacities and durations (3-100hrs) and resulting breakeven costs, and (3) simultaneous optimization based on 11-year weather and load data compared to decision-making based on single-year data.

In summary, the main contributions of this study are: We evaluate Italy's future power system and potential decarbonization pathways including the role of ESS and the impacts of uncertainty in VRE resources; We identify decarbonization scenarios with a wide range of flexibility and low emission technologies including biofuel and CCS; We illustrate the economic viability of different ESS durations up to 100 hours and the evolution of system costs and RES deployment for each ESS duration; We explore the impact of VRE and load data uncertainty on the system's optimal configuration (i.e. installed capacity of generation technologies and their energy share) and costs, and illustrate the consequences of ignoring these long-term uncertainties for supply reliability and operating costs. The rest of paper is structured as follows: section 2 presents the study methodology, section 3 summarizes the inputs and assumption. Case study results are presented in section 4. Finally, section 5 provides discussions followed by conclusions in section 6.

## 2. Methodology

To analyze the role of RES and ESS in the transition towards a decarbonized power system, we conduct a cost-based capacity expansion study. The optimization finds installed capacity and operation of the generation system in year 2050. To perform this analysis, we modified and used GenX, a generation expansion planning (GEP) model [28] with capability of co-optimizing multiple interlinked decision layers of the power system. In this study, the following decision layers are optimized simultaneously within the same optimization problem:

- Capacity expansion (investment decisions of all generation technologies)
- Hourly unit commitment and dispatch of generation and storage units, respecting operational limits
- Scheduling of regulating and operating reserves by thermal, hydro and storage units

GenX is a deterministic model which typically solves capacity expansion problem for one future year. Its objective function minimizes the planning and operation costs of the power system based on the available generation technologies and load requirements as written in (1).

$$\min C = \sum_g \left( C_g^{Inv.} \times A_g \times \delta_g^{Inv.} + C_g^{FixOM} \times \Delta_g \right) \\ + \sum_t \left[ \sum_g (C_g^{VarOM} + C_g^{Fuel}) \times \phi_{g,t} \\ + (C_s^{VarOM} \times \phi_{s,t}) + (C^{ENS} \times \gamma_t^e) \\ + (C^R \times \gamma_t^r) \right] \quad (1)$$

where, $C_g^{Inv.}$ is the investment cost of generation technology $g$

($/MW), $A_g$ is its annuity factor and $\delta_g^{Inv.}$ is its invested capacity (MW). $C_g^{FixOM}$ is the annual fixed operation and maintenance (O&M) cost of generation technology $g$ in $/MW-yr. $\Delta_g$ is the net capacity of $g$ in the target year which is the sum of existing capacity $\delta_g^A$ and $\delta_g^{Inv.}$ minus retired capacity $\delta_g^{Ret.}$ (2).

$$\Delta_g = \delta_g^A + \delta_g^{Inv.} - \delta_g^{Ret.} \qquad (2)$$

$C_g^{VarOM}$ and $C_g^{Fuel}$ are the variable O&M and fuel costs of generation technology $g$ and $\phi_g$ is the energy injected to the system by $g$. $C_s^{VarOM}$ and $\phi_{s,t}$ are the variable O&M cost and charged/discharged energy to/from the ESS. $C^{ENS}$ and $C^R$ are the energy not served (demand curtailment) and unmet reserves penalties, $\gamma_t^e$ and $\gamma_t^r$ are ENS and unmet reserves (MWh) at time $t$. A detailed formulation of the problem in GenX can be found in [28].

To address long-term uncertainty, we are looking at variability of renewable resources and load's historical data across multiple years. In one part of the uncertainty analysis we use single-year data for investment optimization, and then explore the consequences for the system if other weather and load data materialize. Moreover, we also solve the optimization problem across multiple years of VRE and load data to represent these long-term uncertainties within the capacity expansion. Towards this end, we modify the objective function as follows:

$$\min C = \sum_g \left(C_g^{Inv.} \times A_g \times \delta_g^{Inv.} + C_g^{FixOM} \times \Delta_g\right)$$
$$+ \sum_y \left[ W_y \left( \sum_t \left\{ \sum_g (C_{g,y}^{VarOM} + C_{g,y}^{Fuel}) \times \phi_{g,y,t} + (C_s^{VarOM} \times \phi_{s,t}) + (C^{ENS} \times \gamma_{y,t}^e) + (C^R \times \gamma_{y,t}^r) \right\} \right) \right] \qquad (3)$$

The objective function in (3) incorporates multi-year fixed and variable costs and solves the planning and operation optimization for multiple years of weather and load data. $W_y$ is the weight used to normalize the costs to an annual value for year $y$.

The objective functions in (1) and (3) are subject to several groups of constraints including:

*Technology specific:*

(a) Generation units' investment constraints
(b) Thermal, RES and storage units' operational constraints
(c) Unit commitment constraints
(d) Reserves and regulations constraints on individual units

*System level:*

(e) $CO_2$ emission constraints
(f) Hourly energy balance constraint
(g) Total reserves requirements

Group (a) constraints set the limitations on a specific generation technology's capacity, such as maximum hydro power capacity due to resource limitations. Group (b) of the constraints account for the operational limitations of the generation units, such as upper and lower power limits, startup costs and times, resource availability of renewables, and cycling of the storage units. (c) and (d) define the unit commitment and limits on reserve provision from the designated units (thermal, hydro, storage, etc), respectively. Group (e) makes sure that an annual $CO_2$ emission cap (either per MWh or total emissions) are met. (f) assures that the generation and loads are in balance at each hour, unless load curtailment occurs as expressed in (4).

$$\sum_g \phi_{g,y,t} + \phi_{s,t} + \gamma_{y,t}^e + \gamma_{y,t}^r = D_{y,t} \qquad \forall y,t \quad (4)$$

where, $D_{y,t}$ is demand at time $t$ and year $y$. Finally, (g) guarantees the system reserves requirements. The system level reserves are defined as percentage of the load and total VRE capacity. Note that in the formulation with objective function (3), constraint groups (b)-(g) are applied to each individual year. In effect, the stochastic model minimizes the expected cost of meeting demand subject to relevant investment and operational constraints across the years.

The resulting formulation is a mixed integer linear programming (MILP) problem. However, in order to speed up the computational effort, we relax the integrality constraints and therefore solve a linear programming (LP) problem. Earlier studies have indicated that this approximation has limited impact on the generation expansion results [29]. To solve this planning problem, a power system configuration with detailed assumptions for thermal generation units', load data, RES technologies, and energy storage options are required. Also, technology costs and fuel prices' projection for the candidate year (2050) are needed. We present the details of input data and assumptions next.

## 3. Input Data and Assumptions

In recent work, we studied future scenarios of Italy's power system [30], with primary focus on the role of renewable energy sources. Currently, Italy's power system has 35-40% electricity generation from RES. The variation is due to the availability of water resources, as a high percentage of renewable electricity in Italy is harvested from hydro power plants. The results of our work in [30] showed a very high potential for renewables in Italy, with penetration levels possibly exceeding 90% by 2050. However, the study was done with a model without consideration of interannual variability in the VRE. In this paper, we build on the Italy study in [30] to perform an in-depth analysis on the impact of renewable resource variations and also the role of ESS in power system decarbonization.

### 3-1. Generation Units

The current generation technologies in Italy are a combination of conventional thermal power plants (gas, coal and oil) and renewables (hydro, solar PV, onshore wind, geothermal and bioenergy). Based on the decarbonization measures announced by the Italian government in its National Energy Strategy 2017 [31], it aims to phase out coal power plants and significantly reduce oil use by 2025. Therefore, for 2050, we assume that the system will have combined cycle gas turbines (CCGT), open cycle gas turbines (OCGT) and CCGT with carbon capture and storage (CCS) as the conventional thermal unit expansion options. For these units, the fuel prices are collected from the European Commission report on energy, transport and GHG emissions trends to 2050 [32]. The $CO_2$ contents of each fuel was obtained from the US environmental protection agency (US EPA) [33]. Table I presents each fuel's price and its $CO_2$ content. It is important to note that the CCS technology is post-combustion capture at a new CCGT unit with 88% $CO_2$ reduction efficiency [34], and that generation from the biofuel units are assumed carbon emissions-free [35], [36]. Also, minimum output power is considered for the units with unit commitment decisions, which is 30% and 20% of unit size for the CCGT and OCGT plants, respectively. Additional thermal technology parameters include the unit size, heat rate, and start fuel and cost. Thermal units are assumed to have 100% availability. Also, generators' outages and downtime for maintenance are not considered in this study.

For renewable energy technologies, solar PV, onshore and offshore wind, reservoir and run of river hydro plants, geothermal and bioenergy including electricity-only and co-generation units are considered. Among these technologies, hydro power plants are constrained to their resource limitations in Italy which is about 22GW of installed capacity and 45TWh annual energy generation [37]. Reservoir hydro is optimized over the course of the year based on historical water inflows and reservoir capacity limits, whereas run of river hydro is assumed to have a constant average availability for all hours due to the lack of data for hourly variations. The run of river hydro generation is 5.3% of total energy only and therefore, this assumption should not affect the results significantly. Additional input data for reservoir hydro include the initial water level in the reservoir, inflow data, power to energy ratio of the reservoirs, and maximum and minimum reservoir levels. The water level at the end of year is constrained to be within a 10 % deviation from its value in the beginning of year [38]. The installed capacity for hydro plants is the historical data of 2015. No further expansion of hydro is considered in the analysis. The geothermal plants' installed capacity is also considered to remain constant with no expansion possibility. Co-generation bioenergy capacity is also constant, and these plants are modeled as VRE as their main purpose is not electricity generation and therefore, the power output is not schedulable. In contrast, bio electricity-only (from here on referred to as "bio") units are considered to have similar operational characteristics to the thermal power plants with contributions to unit commitment, dispatch, and reserves.

Table II presents the generation technologies for the capacity expansion problem. To account for the variability of the solar and wind powers, their hourly availability factors[1] are obtained from the Renewables Ninja database [39], [40] for 2006-2016 and linearly scaled up to the projected average values of 2050 that are reported in [41]. The resolution of hydro reservoirs' water level is one week. Other technologies have constant availability factors. The investment and fixed and variable O&M costs (Table III) are calculated from European Commission data on technology pathways in decarbonization scenarios [41]. We use a discount rate of 10% to calculate annuities for all technologies. Note that high interest rate is in favor of the technologies with lower CAPEX. Existing capacity of the generation technologies are presented in Table III, too.

Table I. Fuel price projections and $CO_2$ content

| Fuel | Price (Euro/boe) in 2050 | $CO_2$ content (tons/ MMBtu) |
|---|---|---|
| Oil | 111 | 0.070 |
| Coal | 26 | 0.100 |
| Natural gas | 68 | 0.053 |
| Biofuel | 108 | 0 |

Table II. Generation and storage technologies

| Unit | Type | UC*R[†] | Efficiency | Fuel | Average AF[#] |
|---|---|---|---|---|---|
| CCGT | Thermal | UC, R | 0.517 | Gas | 1 |
| OCGT | Thermal | UC, R | 0.341 | Gas | 1 |
| CCGT-CCS | Thermal | UC, R | 0.447 | Gas | 1 |
| Wind: onshore | VRE | - | NA | - | 0.249 |
| Wind: offshore | VRE | - | NA | - | 0.506 |
| Hydro: reservoir | Hydro | UC, R | NA | - | 1 |
| Hydro: run of river | VRE | - | NA | - | 0.445 |
| Hydro: pumped | Storage | R | 0.8 | - | 1 |
| Solar PV | VRE | - | NA | - | 0.173 |
| Battery: 3-hour | Storage | R | 0.85 | - | 1 |
| Battery: 8-hour | Storage | R | 0.85 | - | 1 |
| Geothermal | VRE | - | NA | - | 0.8 |
| Bio electricity-only | Thermal | UC, R | 0.341 | Biofuel | 1 |
| Bio Co-generation | VRE | - | NA | - | 0.54 |

* UC: unit commitment  [†]R: reserves  [#]AF: availability factor

Table III. Investment and O&M costs and existing capacity of generation and storage technologies

| Unit | Inv. cost ($/kW) | Fixed O&M cost ($/kW-yr) | Variable O&M cost ($/MWh) | Existing capacity (MW) |
|---|---|---|---|---|
| CCGT | 748.80 | 17.55 | 2.7 | 0 |
| OCGT | 471.74 | 17.55 | 4.1 | 0 |
| CCGT-CCS | 1755 | 40.13 | 3.25 | 0 |
| Wind: onshore | 1103.31 | 14.04 | 0.21 | 0 |
| Wind: offshore | 2212.47 | 32.76 | 0.46 | 0 |
| Hydro: reservoir | 3510 | 29.84 | 0.37 | 12126 |
| Hydro: run of river | 2691 | 9.48 | 0 | 5332 |
| Hydro: pumped | 4095 | 35.1 | 0.47 | 5732 |
| Solar PV | 531.18 | 10.76 | 0 | 0 |
| Battery: 3-hour | 561.6 | 3.04 | 0.64 | 0 |
| Battery: 8-hour | 971.1 | 3.04 | 0.64 | 0 |
| Geothermal | 3057.21 | 122.85 | 0.37 | 869 |
| Bio electricity-only | 1228.5 | 27.26 | 3 | 0 |
| Bio Co-generation | 1228.5 | 27.26 | 3 | 2040 |

---

[1] i.e. available energy before any potential curtailment.

### 3-2. Energy Storage Options

We assume that the installed pumped-hydro storage capacity has a duration of 8 hours and that this capacity remains constant. We consider two electrochemical storage technologies as expansion alternative, with 3 and 8 hours duration, respectively. Storage systems and thermal units contribute to providing regulating and operating reserves. Investment and O&M costs for the battery energy storage are calculated from [42]. The main assumptions for energy storage are summarized in Table II and Table IIITable II.

### 3-3. Load and System Representation

To predict the hourly load variations for 2050, the historical load data of 2006-2016 are scaled up based on the total load growth projections [30], i.e. the average load equals the projected value and the hourly variations are captured from the historical data. In these calculations, we assume that the Italian power system has constant 7% transmission and distribution losses which is added to the projected load and there will be no electricity import or export in 2050 which was 35 TWh in 2015. The latter assumption is motivated from a national energy security perspective. Based on these calculations, total electricity generation requirement in Italy will be 404 TWh.

The main purpose of this study is capacity expansion analysis and therefore, the transmission network limitations are not reflected in the optimization. To reduce computation time, power plants are aggregated and clustered into groups, where the plants in each group have identical characteristics. We assume that the system's operating reserve requirements are a function of the hourly load and VRE generation, i.e. 3% of load and 5% of VRE. Frequency regulation reserve requirements are set to 1% of the hourly load [43]. Finally, the value of lost load is set to $13,000 per MWh [44] and the unmet reserves penalty is $1000 per MW-h.

### 4. Results

This section presents results of different simulated scenarios to quantify the impact of energy storage and weather and load data uncertainties on the low or zero $CO_2$ emission system's configuration, operation and costs. The simulated scenarios can be categorized as follows:

- $CO_2$ emission constraints (zero to no-limit) with and without battery energy storage (3-hour and 8-hour) investment options for different weather and load data (2006-2016)
- Energy to power ratio (duration) of energy storage (3-hour to 100-hour) combined with different fixed capacities of energy storage (1, 10 and 100GWh). The cases are run for different weather and load data (2006-2016) with a zero $CO_2$ emission limit.
- Operation of system in different years (2006-2016) in case of deterministic decision-making based on single year data (2006-2016) or based on stochastic optimization over 11 years. These cases are run with battery energy storage (3-hour and 8-hour) and zero $CO_2$ emissions limit.

### 4.1. $CO_2$ Emission Constraints and Battery Installation

In this section, we first optimize the power system with no constraint on the $CO_2$ emissions and calculate the resulting emissions under the resulting optimal expansion and operation. Then, we impose several annual $CO_2$ constraints on the model (100%, 50%, 10%, 5% and 0% of the unconstrained value). Furthermore, in order to investigate the impact of energy storage on the system expansion plan, we do the analysis with and without the battery (3-hour and 8-hour) investment options.

#### 4.1.1. Power system configuration

The results show a high RES penetration in the optimal power system configuration even without $CO_2$ constraints (Figs. 1-2). The RES investments are driven by projected reductions in RES technology costs. Also, the presence of battery energy storage contributes to increase the RES penetration level. Without any $CO_2$ constraint and on average across the results of 11 years (2006-2016) single optimizations, without batteries, the RES installed capacity is 80% in 2050, which compares to 45% in 2015. Moreover, RES meets 73% of the electricity demand (compared to 34% in 2015). When the batteries are introduced to the system, the RES capacity increases to 91% with an 85% share of the annual generation.

Fig. 1 shows the energy contribution of each generation technology to the total demand for different $CO_2$ constraints. The energy shares vary significantly across weather years regardless of the $CO_2$ constraint, as indicated by the error bars. The figure also reveals the impact of battery energy storage installation on the annual generation, and its variability, of each technology. Without battery, the absolute deviation (max-min) from the average value across the 11 years for gas power plants is 13%, while it is 19% for wind power, 9% for solar power and 20% for bio. With battery installations, the impact of the weather and load uncertainties become larger. In this case, gas power plants, wind energy, solar and bio have 20%, 35%, 11% and 40% variation in their energy shares, respectively. This result indicates that the uncertainty in energy contributions increases by adding more storage. This is because high storage installations lead to increased adoption of VRE and as a result increased annual variation causing higher uncertainty in the results. This finding emphasizes the importance of considering weather and load variations in the decision-making process.

Comparing the annual generation of each technology across $CO_2$ scenarios shows that imposing higher $CO_2$ constraints leads to a transition from gas power plants to CCS and then bio for flexibility. The CCS is slightly cheaper than bio, however, its carbon capture efficiency is not 100% and therefore it is not suitable for a zero-emission system. Tighter $CO_2$ constraints increases the share of RES. Adding battery energy storage to the system has two significant impacts compared to the system without battery. First, in presence of the batteries, the need for flexible generators (gas and bio) decreases as flexibility is provided by the batteries. Second, the share of solar energy increases substantially, while the share of wind energy decreases, indicating that energy storage from batteries promote solar energy. Note that, as expected, we do not observe noticeable changes in the hydro utilization (including reservoir and river) due to its capacity which is constrained to the present level. These results are also reflected in the installed capacity of each technology. Fig. 2 shows that the total installed capacity of the flexible generators is almost constant under different $CO_2$ constraints, but there is a transition from regular gas to CCS and then bio under decarbonization. Also, the installed capacity of renewables increases with tighter $CO_2$ limits, as expected. However, without batteries, this increase happens primarily in wind energy. In contrast, with batteries the increase happens primarily in solar energy. Increased installed capacity of solar and wind energy in different $CO_2$ and battery scenarios affects their annual average curtailment calculated as in (5) and shown in Fig. 3.

$$\Theta_{g,y} = \frac{\sum_t \phi_{g,y,t}}{\sum_t F_{g,y,t} \times \Delta_g} \quad (5)$$

where, $F_{g,y,t}$ is the availability factor of generation technology $g$ at time $t$ and year $y$. Based on this figure, solar curtailment is significantly lower than the wind curtailment (on average across simulated $CO_2$ scenarios, 4% compared to 23%). The capacity expansion model logically curtails the wind before the solar, because of the wind's higher variable O&M cost. Introducing tighter $CO_2$ constraints increases the percentage curtailment of renewables, due to their higher installed capacities. Note that the solar curtailment percentage is higher in the case with battery than in the case without. This is because the installed capacity of solar is significantly higher when batteries are part of the resource mix. For the wind energy, the curtailment percentage ins lower with battery, due to the lower installed capacity. Another important finding is that a battery duration of 8 hours is preferred over 3 hours. The installed capacity of 8-hour batteries is 2-5 times higher than that of the 3-hour battery in different $CO_2$ constraints. Table IV summarizes the variation in energy and capacity shares by technology across 11 years of weather and load data. The variations are substantial, particularly in the zero emissions systems. Note that the technologies not included in the table had insignificant variations across the 11 scenarios.

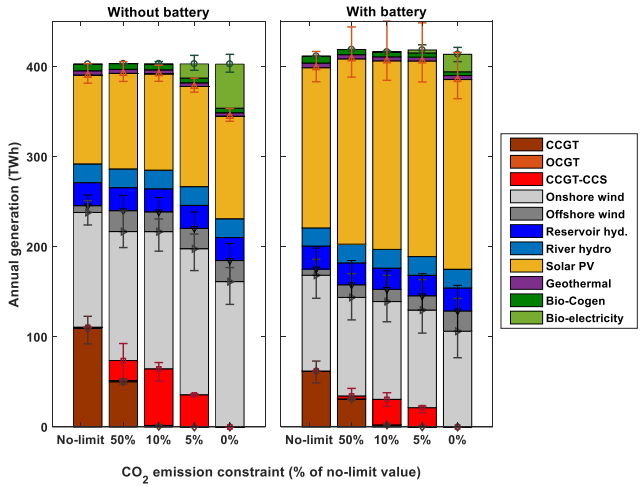

Fig. 1. Annual generation of technologies under different $CO_2$ constraints, with and without battery (the error bars show variation over 11 years of individual optimization results due to the weather and load data differences)

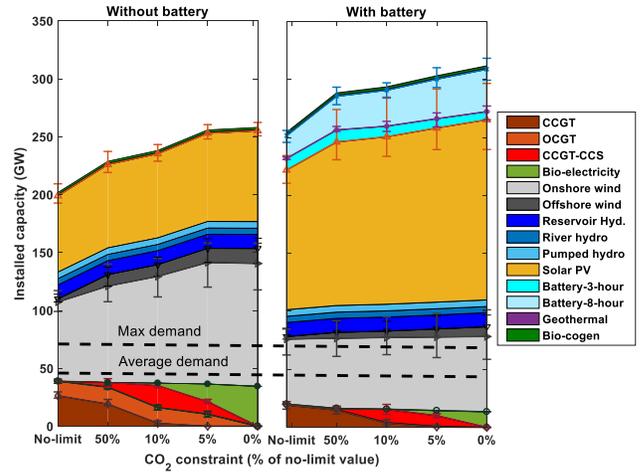

Fig. 2. Installed capacity of generation technologies under different $CO_2$ constraints with and without battery (the error bars show variation over 11 years of individual optimization results due to the weather and load data differences)

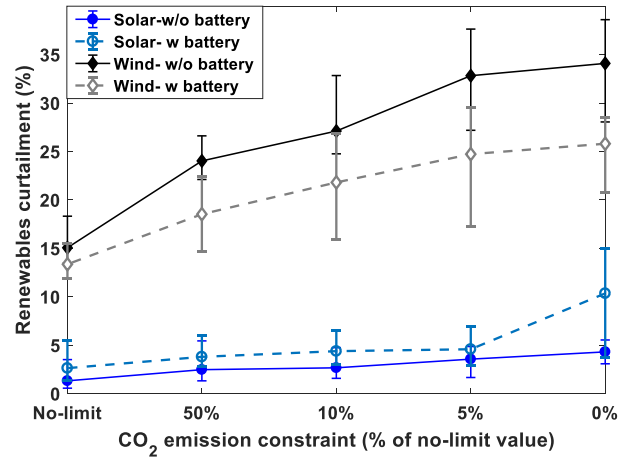

Fig. 3. Solar and wind curtailments in different $CO_2$ and storage scenarios (error bars show variation over 11 years of individual optimization results due to the weather and load data differences)

Table IV. Variation in annual generation and installed capacity over 11 years of weather and load data (inside the brackets: minimum, average and maximum values, respectively)

|  |  | Without battery | | With battery | |
|---|---|---|---|---|---|
|  |  | Without $CO_2$ limit | Zero emission | Without $CO_2$ limit | Zero emission |
| Energy (TWh) | Gas | [92 110 124] | 0 | [49 62 74] | 0 |
|  | Wind | [114 135 161] | [136 185 219] | [80 113 154] | [76 128 143] |
|  | Solar | [90 100 111] | [108 115 123] | [162 177 196] | [189 210 240] |
|  | Bio | 0 | [40 50 60] | 0 | [11 20 27] |
| Capacity (GW) | Gas | [35 40 44] | 0 | [15 20 22] | 0 |
|  | Wind | [60 70 85] | [83 120 145] | [42 68 75] | [45 73 106] |
|  | Solar | [60 66 76] | [74 78 86] | [109 120 133] | [130 155 187] |
|  | Bio | 0 | [31 35 38] | 0 | [8 13 19] |

*4.1.2. System investment and operational costs*

Including battery energy storage in the power system decreases the need for thermal power plants and as a result reduces the $CO_2$ emissions and lowers the system's total costs. On average across 11 years data, $CO_2$ emissions per kWh consumed electricity decrease from 97g to 54g in the no-limit case. It is worth noting that $CO_2$ emissions per kWh in 2015 was 275.5g. The average energy cost in a year is calculated by dividing total system cost by total load and shown in Fig. 4.

As shown in this figure, the average energy cost per MWh decreases with battery installations for all $CO_2$ emission constraints, under the assumed technology and fuel costs (Tables I and II). The figure also indicates that, in the absence of a carbon constraint, the total $CO_2$ emission is reduced by 45% on average. This is due to the lower share of gas power plants when batteries are part of the resource mix (Fig. 1). Uncertainties in the weather and load profiles change the total system cost and as a result the average energy cost significantly. The impact of these uncertainties is larger with higher constraints on $CO_2$ emissions, due to the higher penetration of renewables. In the zero-emission scenario, the yearly average energy cost (y-axis) varies 10% between different weather and load data, while the variations are smaller in the unconstrained scenario due to lower amounts of VRE. However, note that the variation in the $CO_2$ emissions due to these uncertainties is considerable in the no-limit scenario (40%, as indicated by the x error bars on this point in the figure).

Exploring the energy cost breakdown (Fig. 5) shows that almost 20% reduction in energy cost can be achieved with batteries in the zero-emission power system. However, even without any $CO_2$ constraint, batteries can lead to 7% cost reduction for the given assumptions. The dominant costs in the system are the fixed costs. Variable cost decreases with lower $CO_2$ constraints due to lower use of fuel in the gas power plants. However, it increases in the zero-emissions case, because of a switch to bio-fired plants with high variable costs. (CCS is not part of the solution in the zero-emission case since the capture rate is less than 100 %,) Although the energy not-served (ENS) and unmet reserves costs are very low compared to the fixed and variable costs, the detailed results show that ENS is zero with batteries, while it can go up to 20GWh (0.005% of total load) without batteries. The unmet reserves of 40-60 GW-h without batteries are decreased to 0.7-1.7 GW-h with batteries. These results indicate the potential of batteries to enhance the reliability of electricity supply while reducing the energy cost, if the projected cost reductions for energy storage technologies are realized.

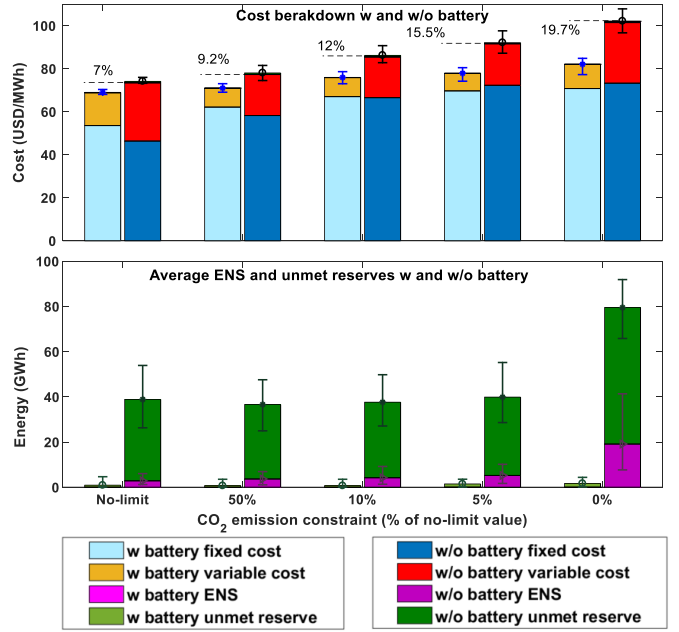

Fig. 5. Energy cost breakdown (upper trace) and ENS and unmet reserves (lower trace) for different $CO_2$ constraints with and without battery

### 4.2. ESS Energy to Power Ratio (Duration)

In this section, we explore the impact of the ESS duration on the power system costs and renewables adoption for the zero-emission case. For more detailed analysis of this dimension, we defined ESS scenarios with three energy capacities (1, 10, and 100 GWh) and seven durations (3, 8, 15, 24, 48, and 100 hours) and run these scenarios for all weather and load years. In each capacity scenario, the ESS energy capacity is kept constant and its power is changed to reflect the intended storage duration. Hence, the amount of ESS is an exogenous input and corresponding investment costs are not considered. The model optimizes the rest of the system for the defined ESS capacity, then we calculate the breakeven investment cost of ESS, $c^{BE}$, per unit of storage using (6), where $r$ is interest rate, $\Delta_{ESS}$ is the energy capacity of the ESS and $L$ is the ESS lifetime in years which is considered 25 years in this study.

$$c^{BE} = \frac{\left|C_y^{tot,noESS} - C_y^{tot,ESS}\right| \times (1 - (1+r)^{-L})}{r \times \Delta_{ESS}} \quad (6)$$

Fig. 6 presents the system's total costs without ESS investment cost and the breakeven cost of ESS for different capacities and durations. The total system saving (relative to the case without ESS) is higher for larger ESS capacities, as expected. Also, the saving from ESS is larger for shorter storage durations of the ESS. This is because the energy capacity is constant across storage durations, so that a shorter duration battery has larger power capacity. We calculate the breakeven cost for the ESS from the total system cost reduction. The results in Fig. 6 indicate that regardless of the ESS capacity, the breakeven cost is very low for long ESS duration (high E/P ratios), reaching 85 USD/kWh for the 100-hour duration. Note that this cost includes the balance of system (AC/DC conversion) costs. However, for shorter durations, there is significant difference in breakeven cost between ESS capacity scenarios, i.e., the lower the ESS capacity, the higher the breakeven cost, indicating that the first installations of ESS are the most valuable. Comparing the breakeven costs to projected Li-ion battery costs bounds in 2050 [42] indicates that Li-ion batteries may become economically viable for shorter durations first, however for the very long durations, the breakeven cost is in the lower range of the projected future Li-ion battery costs. Hence, our results indicate that long-duration storage is not critical for decarbonization of the system considered in this case study and that technology breakthroughs would be needed for such technologies to become viable compared to other alternatives. Note that we assumed a storage efficiency of 85% in these analyses, which is significantly higher than what is the case for long duration storage technologies today, such as hydrogen. Exploring the effect of weather and load data uncertainties shows that the variation in breakeven cost is higher for shorter durations and lower ESS capacities. For 100-hour ESS, the effect of data uncertainties is less than 5% for all ESS capacities. For 3-hour ESS, the deviations are 30%, 20% and 12% for 1GWh, 10GWh and 100GWh ESS, respectively.

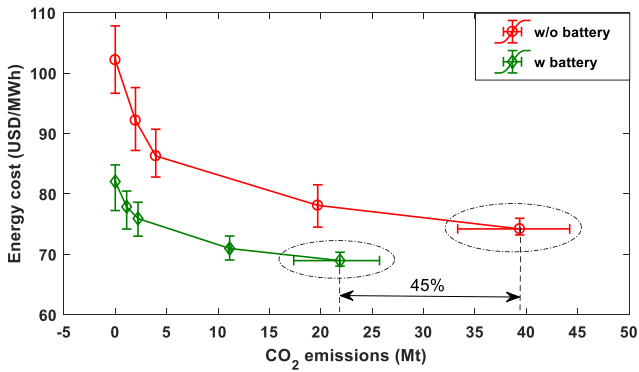

Fig. 4. Energy cost per MWh vs $CO_2$ emissions with and without battery and their variations across different weather and load data

Different ESS durations trigger installation of different types of renewables. Fig. 7 shows the annual generation for wind, solar and bio plants across different ESS capacities and durations. Without ESS, the system is dominated by wind power. Introducing lower duration ESSs leads to high adoption of solar energy and therefore decrease in the wind and bio shares. In particular, 1-hour and 3-hour ESSs significantly boost the solar contribution This can be explained by the fact that solar generation is in its peak for several hours at midday, requiring short duration ESS to shift these hours' generated energy to the evening peak load hours. With longer duration storage, the energy contributions from the different technologies gradually revert to their levels without energy storage (y-axis intersect). Note that at the largest storage installation level (100GWh), the share of the bio does not go back to the level of without ESS, because ESS provides the required flexibility instead of bio.

### 4.3. Optimizing over Single vs. Multiple Years of VRE and Load Data

The previous sections illustrated that uncertainties in the weather and load data can significantly change the system's optimal configuration. However, the optimal solution based on a single year's data may lead to suboptimal operation in other years due to these uncertainties. To better account for the weather and load uncertainties in the decision-making process, we analyzed a case where the model in equation (3) optimizes the expansion build-out over all the 11 years of weather and load data based on expected costs.

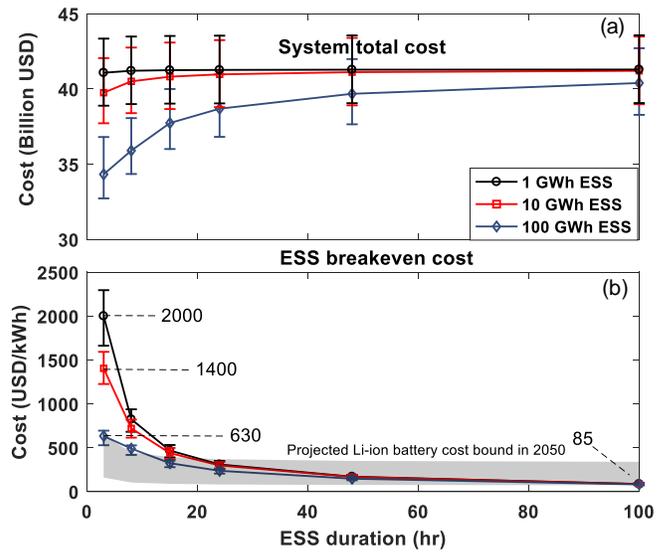

Fig. 6. (a) Total cost and (b) ESS breakeven cost for different ESS capacity and duration scenarios for zero-carbon system. Error bars across 11 years of data.

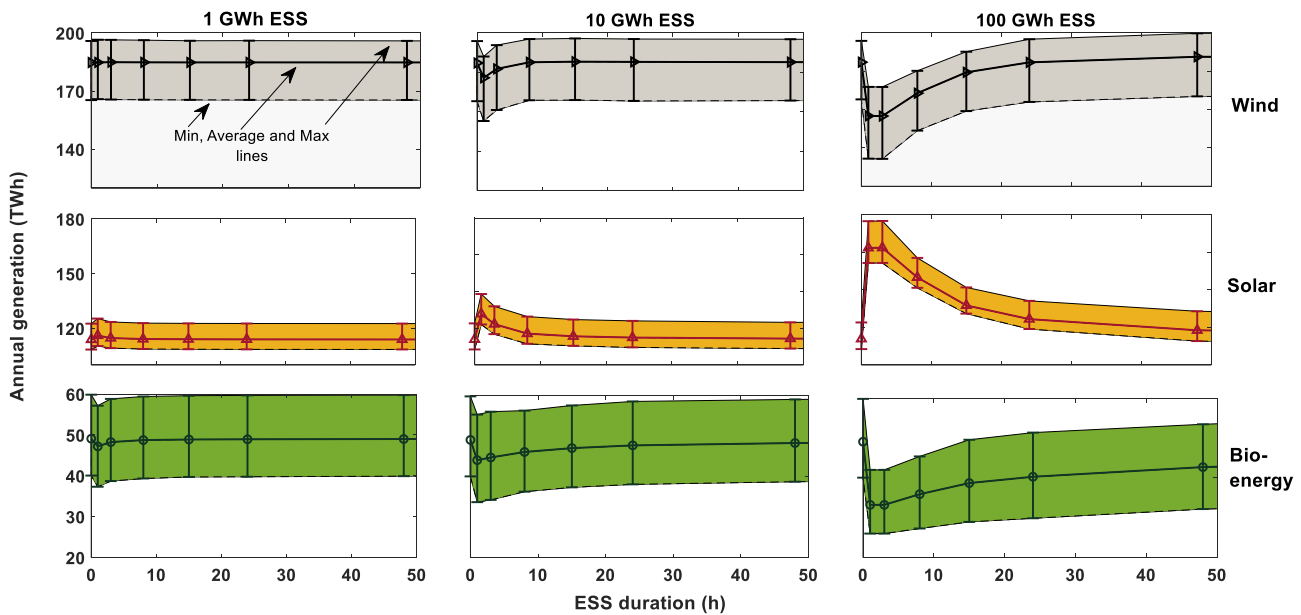

Fig. 7. Annual generation from wind, solar and bio in different ESS capacity and duration scenarios

Consequently, we compared the results to the optimal system portfolio based on single-year data from the model in (1). All these optimizations are for the zero $CO_2$ emission case. It important to note that the average annual availability factor of VRE in different years data are scaled up to the same value for 2050, and only the hourly variation pattern is adopted from different historical year-data (2006-2016).

Fig. 8 shows the variation of installed capacity and annual generation of renewables in different single year solutions and for the case that optimizes across 11-years by minimizing expected costs. Solar PV's installed capacity has the minimum variation across years among renewables with ±20% deviation from the average, while offshore wind variation can go up to ±100%. Installed capacities for onshore wind and bio vary ±30% and ±40% from their average values, respectively. These variations in the installed capacity lead to similar variations in each technology's energy share. Also, comparing the results of 11-year versus single-year optimization indicates that although the 11-year results are always in the bound of single-year outputs, the optimal solution for 11-year co-optimization (i.e. the green dot) is different from the average of the single-year optimizations (red dots).

Considering the variations observed in the results using different weather and load data, we examined what happens in other years if the system planning decision is made based on single year data. For this purpose, the optimal installed capacities for one year is used as fixed input for the other 10 years and the system's operation is optimized with those fixed capacities. This process is also repeated for the optimal capacities of the 11-year optimization case. The results reveal substantial differences in the system's operating conditions and corresponding costs in different weather/load years, as shown in Fig. 9. The system's investment costs in different years do not vary significantly (59-63 $/MWh) and therefore, it is possible to compare the system's performance in different years based on the operating costs. With our data set, basing expansion decisions on years 2009 and 2010 give the worst performance, i.e. the system costs in other years can go extremely high due to high variation between the VRE output and the load, and resource inadequacy. On the other hand, the 2008 and 2015 results show that if the system is designed based on data for these years, it will also perform well in the other 10 years and the variation in the system operational costs is very narrow. This is driven by a string of low VRE resources and high load times for 2008 and 2015 and the system compensates the lower VRE generation with higher capacity installations. In fact, these two years' results are similar to the result for the 11-years case, which optimizes for all years together.

The rest of the years' results are not as unfavorable as 2009 and 2010. However, they have one or two outlier points which shows an important finding: even though decision-making based on those years' data will lead to proper operation of the system in most other years, there will be one or two years where the system will operate with unreasonably high cost. The source of high operating costs is illustrated in the cost breakdown plot in Fig. 10. The figure shows that when an unsuitable weather and load data year (2009) is used for decision-making, there will be capacity inadequacy and high load curtailments under more than half of the weather/load years, which is very costly for system operations on average. In contrast, the system optimized based on 2015 data has very low load curtailment costs in other years, similar to the system based on 11-year optimization. This result highlights the high sensitivity to decision-making base year under single-year planning. A robust capacity expansion plan can be achieved by selecting the worst-case scenario for the weather and VRE output which leads to a generation technology mix with better performance in the rest of the years in our case study. As the average annual availability factor of VRE in different years are the same and only the hourly variation pattern is different, the difference between good and bad base weather-year is a function of the synchronization between VRE and load data. In hours with high load and low VRE output, load curtailment happens which increases the ENS cost and as a result total operating cost. Overall, our results indicate that generation expansion analysis should be conducted across multiple weather and load years, whenever this is possible.

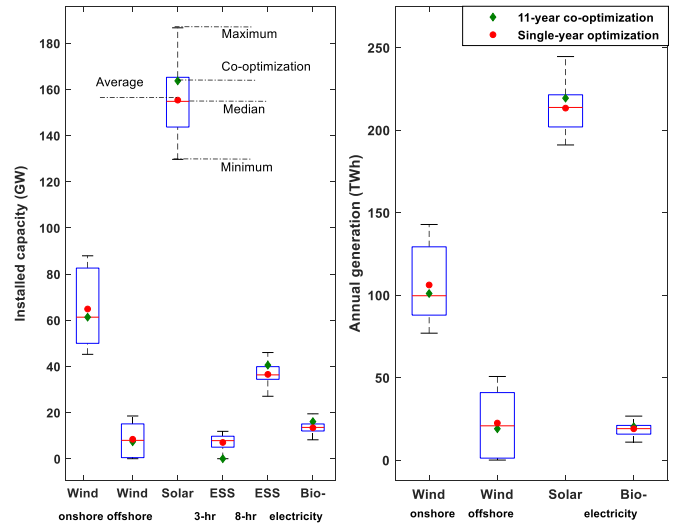

Fig. 8. Installed capacity and annual generation for selected technologies: comparison of single-year and 11-year optimization

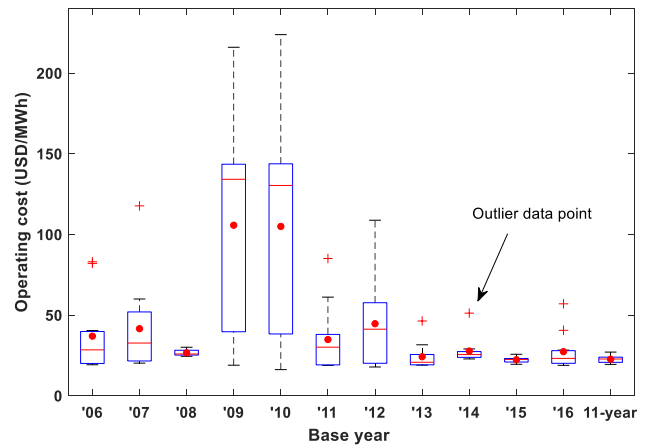

Fig. 9. System's operating costs variation in different years with decision-making based on single-year and 11-year data

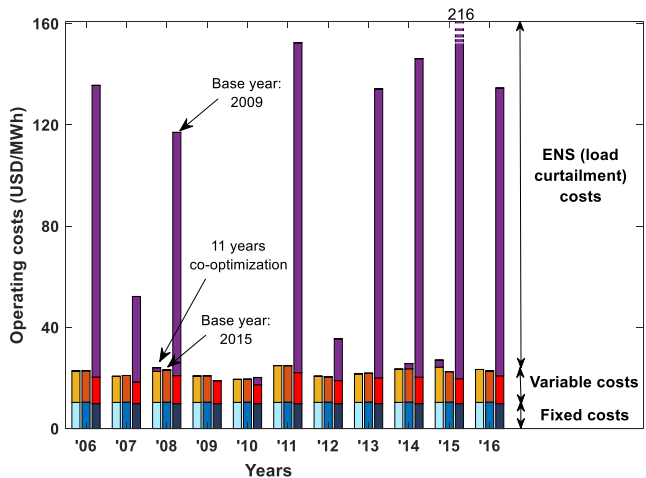

Fig. 10. Cost breakdown for the base years with lowest (2015) and highest (2009) average operating costs and 11-year optimization

## 5. Discussion

Renewable energy sources play a critical role in the pathway to zero-carbon power system. However, due to their high variability, without ESS, the optimal generation expansion plan tends to employ more expensive flexibility options such as bioenergy and natural gas with CCS. ESS enables higher deployment of RES sources such as solar and decreases the overall system investment and operational costs. Mass deployment of RES and ESS are economically viable if the projected technology cost reductions are realized and there is enough natural potential for RES and ESS resources and no social or political limitations on their expansion. These limitations are not modeled in this study. Also, this study does not consider the network configuration, transmission and distribution systems limitations. The results should be interpreted accordingly.

Our results indicate that a zero-emission power system in Italy can be achieved with lower costs by using energy storage. Full decarbonization without energy storage increases the energy generation cost by 40%, while the cost increase can be reduced to 20% with deployment of batteries. This cost difference is mostly due to the higher share of bio-generation in the absence of batteries, as bio-plants are an expensive provider of flexibility. The resource potential for bio-energy generation should also be explored in more detail. Note that the increase in system cost from decarbonization is higher for stricter $CO_2$ emission constraints, e.g. moving from no-limit to 10% emissions has a lower rate of cost increase than cutting the last 10% of $CO_2$ emissions. Therefore, getting to zero-carbon power system is considerably more costly than getting to 5 or 10% emissions, which is in line with other recent findings [45]. Our study also shows that the load and reserves curtailment significantly decrease with ESS installations, indicating that ESS may enhance the reliability of electricity supply in a cost-effective manner. With their high efficiency, batteries provide stored cheap renewable energy in peak hours to prevent demand curtailments. They also have high flexibility and short response times, thereby replacing the need for slower and costlier generators to provide reserves.

The added value of ESS to the power system planning and operation depends on its energy capacity and energy to power ratio (duration). ESS with higher energy capacity leads to more decrease in the system costs. However, the added value of ESS does not increase linearly with its installed capacity, as the marginal benefits of ESS are declining for higher levels of installed capacity, in line with the findings in [17]. Also, among the simulated durations, ESS with durations shorter than 24 hours have the most significant impacts on the system configuration and costs. This is because of daily demand and VRE (wind and solar) output cycles. Among these ESS durations, shorter ESS promotes solar energy, while longer ESS leads to higher wind energy deployment. We have not analyzed EES durations beyond 100 hours, i.e. seasonal energy storage systems in this study.

Another important issue that arise while making capacity expansion decisions is the uncertainties in the VRE output and demand data due to the annual weather variations. Our analysis of 11 years of VRE and load variations indicated that the data uncertainties can significantly alter the optimal expansion decisions and it is crucial to consider these uncertainties in the planning analysis. Decision making based on single year data, depending on the hourly VRE and demand variations, may lead to higher system costs in other years. In this analysis, we only considered the impacts of different hourly VRE availability and load while keeping the average annual capacity factors constant. We still found substantial differences in the operation of system and the optimal capacity expansion. Therefore, if a year with high coincidence between VRE outputs and demand is used for the planning optimization, the system will experience shortfalls in years with lower VRE output and load matching. If the uncertainty of total load in a target year and total VRE generation is added to the uncertainty of hourly variations, the consequences will be even worse. Therefore, in order to find a robust solution on the capacity expansion problem, planners should consider multiple years of VRE and load data in the optimization. Alternatively, using a worst-case scenario of VRE and load in single year planning will also provide a more robust expansion plan.

Finally, it is important to note that although the assumptions and inputs are carefully defined and obtained from credible sources, this study is performed under a set of assumptions that are highly uncertain, particularly considering the long horizon of 2050.

## 6. Conclusion

In this paper, we explored the decarbonization of electricity supply with a particular focus on the impact of ESS duration and variability in the hourly VRE and load data in a case study for Italy. Our results show a high penetration of RES even without a $CO_2$ constraint or economic RES incentives driven by projected technology cost reductions. The penetration of RES increases further under decarbonization constraints. Our results indicate that gas power plants with CCS will play a minor role in decarbonization due to carbon capture efficiency less than 100 %.

Under the given technology cost assumptions, system planning and operation is less expensive with ESS, since it enables higher deployment of cheap renewable energy technologies. Although ESS increases the electricity supply reliability by reducing the ENS, the impact of the variability in weather and load patterns become larger due to more VRE in the resource portfolio. The solar curtailment also increases with ESS in the optimal generation mix due to the higher investments triggered by ESS. Another important observation is that long duration ESS (10-100 hours) only have a modest marginal value per kWh, while shorter duration ESS (less than 10 hours) address fluctuations in renewables on a daily basis and add significant value to the system. Also, the first installed capacities of ESS leads to higher saving in the system costs, and therefore higher breakeven cost.

Variations in the hourly patterns of VRE and load bring significant uncertainty to the system configuration and costs. Our results from the case study of Italy indicate that energy costs can vary up to 10% from year to year based on the individual year that the system is optimized for. Moreover, decision making based on one-year of VRE and load data can easily lead to suboptimal operations in other years, causing unreasonably high load curtailments and a substantial increase in the energy costs. Overall, our results illustrate that the best solution is to perform capacity expansion optimization across multiple years of weather and load data.

Directions for future work include optimization of ESS duration within the capacity expansion, modeling the impact of ESS degradation on optimal portfolio planning, conducting additional sensitivity analysis for a wider range of technology cost and fuel price assumptions, and factoring the transmission network and grid stability into the analysis.

**Acknowledgement**